\documentclass[aps,pre,twocolumn,superscriptaddress,floatfix,showkeys,a4paper]{revtex4}

\usepackage[dvips]{graphics}
\usepackage{amsfonts}
\usepackage{pgf}

\usepackage{color}

\newcommand{\jhpd}[1]{#1}
\newcommand{\jhpdtwo}[1]{#1}
\newcommand{\HS}[1]{#1}

\newcommand\ba{\begin{eqnarray}}
\newcommand\ea{\end{eqnarray}}
\newcommand\bc{\begin{center}}
\newcommand\ec{\end{center}}
\newcommand\ra{\rightarrow}
\def\nn{\nonumber}

\def\epsilon{\varepsilon}
\def\e{\ensuremath{\mathrm{e}}}
\def\p{\ensuremath{\partial}}
\def\calD{\ensuremath{\mathcal{D}}}
\def\calL{\ensuremath{\mathcal{L}}}
\def\calLeff{\ensuremath{\mathcal{L}_{\mathrm{eff}}}}
\def\R{\ensuremath{\mathbb{R}}}
\def\ra{\rightarrow}

\newcommand{\bi}{\begin{itemize}}
\newcommand{\ei}{\end{itemize}}
\newcommand{\be}{\begin{enumerate}}
\newcommand{\ee}{\end{enumerate}}

\begin{document}

\title{Variational approximation and the use of collective coordinates}
\author{J.H.P. Dawes}
\affiliation{Department of Mathematical Sciences, University of Bath, Claverton Down, Bath BA2 7AY, UK}
\author{H.\ Susanto}
\affiliation{School of Mathematical Sciences, University of Nottingham, University Park, Nottingham, NG7 2RD, UK}


\begin{abstract}
We consider propagating, spatially localised waves in a class of equations that
contain variational and non-variational terms.
The dynamics of the waves is analysed through a collective coordinate approach.
Motivated by the variational approximation, we show that there is a
natural choice of projection onto collective variables
for reducing the governing (nonlinear) partial differential equation (PDE) to coupled
ordinary differential equations (ODEs). This projection produces ODEs whose
solutions are exactly the stationary states of the effective Lagrangian that would be
considered from applying the variational approximation method.
We illustrate our approach by applying it to a modified Fisher
equation for a travelling front, containing a non-constant-coefficient nonlinear term.
We present numerical results that show that our proposed projection captures both the
equilibria and the
dynamics of the PDE much more closely than previously proposed projections.
\end{abstract}

\maketitle

\section{Introduction}
\label{sec:intro}

The dynamics of localised waves in completely integrable systems can be analysed notably using the inverse scattering, or inverse spectral, transform. The method was first proposed in \cite{gard67} to solve the Korteweg--de Vries (KdV) equation \cite{bous77,kort95}. It provides an explicit integral representation of the solution that has then been applied to many nonlinear evolution equations, such as the sine-Gordon equation \cite{ablo73,ablo74} and the nonlinear Schr\"odinger equation \cite{zakh72,zakh73} (see, e.g., \cite{wang02} for a list of other `integrable' partial differential equations (PDEs)). Localised waves of integrable systems are also often referred to as solitons \cite{zabu65}. Initially studied in autonomous systems, the concept of integrability \jhpd{and the notion of
`soliton'} has been extended to nonautonomous equations \cite{calo76,chen76,serk07}. 

Even though exact integrability plays an important role in the theory of nonlinear waves, most equations modelling physical systems are not integrable. \jhpd{For systems that are `close' to integrable, the inverse scattering transform
method} can still be employed. By treating the terms that cause nonintegrability as a perturbation, an inverse scattering perturbation theory was introduced in \cite{karp77,karp77a,kaup76}. See \cite{kivs89} for a comprehensive review of the method and its applications. The main limitation of the \jhpd{perturbation theory is, naturally, that} it relies on the integrability of the unperturbed system. 

A different approach is provided by a so-called collective variable or \jhpd{collective} coordinate method. By allowing \jhpd{the} parameters of the localised wave, such as its position or width, to vary as a function of time, one then reduces the solution dynamics from being governed by a PDE into coupled ordinary differential equations (ODEs). The reduction \jhpd{can be obtained in several different} ways, for instance, by exploiting conservation laws of the unperturbed equations or Lagrangian formalisms. Thus, instead of integrability, these direct perturbation methods require conserved quantities.
In the first of these methods, also called energetic analysis, the dynamics of the solution's parameters are governed by the rate of change of the integral invariants, which should vanish in the unperturbed case (see, e.g., \cite{berg83,mcla78,pere77,kivs89}). In the Lagrangian or variational method, the solution's parameters act as new canonical coordinates that have to satisfy the Euler-Lagrange equations (see, e.g., \cite{rice80,rice83,M02,CP11}). The estimation of optimal parameters through the construction of variational principles has a long history, scattered across
many different application areas, including for example fluid mechanics where Chandrasekhar \cite{C61} used a variational
approach to estimate the critical Rayleigh number for the onset of thermally-driven instability in a
layer of viscous fluid.
We refer the reader to \jhpd{ref.} \cite{sanc98} for a review of collective coordinate approaches.

Recently, Caputo and Sarels \cite{CS11} considered the interaction of a reaction-diffusion front and a localised defect. A theoretical analysis based on collective coordinate approach was carried out. Instead of using the Lagrangian formulation, Caputo and Sarels derived equations of motion for the parameters by using conservation quantity-like integrals, which were obtained by substituting the ansatz for the solution into the PDE and averaging it after multiplications with a function (see also \cite{ott69,boes88,boes89} for a similar approach). However, it is important to note that the multiplying functions \jhpd{were, in that study,} chosen rather arbitrarily. Here, we show that there is a natural choice of functions for projection onto the collective coordinates, based on the variational formulation. Our proposed method is similar to that of \cite{ande78,bond79,chav98,kivs95} when the non-variational terms are small. \HS{It is also similar to the so-called nonlinear Ritz method for dissipative systems \cite{abra08} (see also the Supplemental Material of \cite{vaku09}), which is a generalization of the Whitham principle for periodic waves \cite{whit74}. The Ritz principle separates governing equations into a variational functional that defines a free energy and a functional that depends on the time derivative only. Our proposed method extends the principle to the case when the free energy function may also contain non-variational terms. 
} 

We then apply our proposed method to a modified Fisher equation, i.e.\ that considered in \cite{CS11}. The reader is referred to \cite{CS11} and references therein for the relevance and the physical applications of the equations. We \jhpd{show below} that our method, which in the present example coincides with that in \cite{abra08}, provides a \jhpd{very} good approximation to the numerical results. Due to the inhomogeneity, the front can be pinned. The pinning estimate will never be excellent because the front departs from the form of our ansatz as it crosses the inhomogeneity. Radiation of continuous waves is also disregarded as a typical characteristic of collective variable approaches. The ansatz we use is expected to apply when the inhomogeneity is slowly-varying, but not otherwise. Nevertheless, notably we observe that even when the inhomogeneity is a $\delta$-function, we obtain \jhpd{good quantitative agreement between the approximation and numerical solution of the full problem.}

The remainder of the paper is structured as follows.
In section~\ref{sec:dyn} we write out the general expressions that
indicate how one can extend the variational approximation into
a set of dynamical equations for the time evolution of collective
coordinates. In section~\ref{sec:fisher} we apply our method
to a specific problem: the propagation of fronts in a spatially
inhomogeneous medium. In section~\ref{numr} we show that our results
here on the dynamics of the problem are a substantial improvement over previous work,
comparing them with numerical simulation and the results of \cite{CS11}.
A summary and concluding remarks are contained in section~\ref{sec:discussion}.

\section{Dynamics and the variational approximation}
\label{sec:dyn}

Let $u(x,t)$, for $x \in \Omega \subseteq \R$ and $t \geq 0$,
satisfy the PDE
\ba
u_t & = & f_V(x,u,u_x,u_{xx}) + f_{NV}(x,u,u_x,u_{xx}), \label{eqn:u}
\ea
which is second-order and (in general) inhomogeneous in space.
\jhpd{We note in passing that our results extend naturally to PDEs of
higher order in space, but~(\ref{eqn:u}) is sufficient here to describe our
approach.} We assume
that suitable boundary conditions are specified on $u|_{\p \Omega}$, for
example Neumann boundary conditions, $u_x=0$. \jhpd{In writing~(\ref{eqn:u}) we
have split} the right-hand side into variational and non-variational parts, i.e.
there exists a function $F(x,u,u_x)$ such that
\ba
f_V(x,u,u_x,u_{xx}) & = & \frac{\p F}{\p u}
- \frac{d}{dx}\frac{\p F}{\p u_x}. \label{eqn:fv}
\ea
In the variational case (that is, assuming that $f_{NV}\equiv 0$)
we observe that stationary solutions of~(\ref{eqn:u}) are extrema
of the functional $\calL:=\int_{\Omega} -F(x,u,u_x) \ dx$ taken over
a space of suitable functions $\calD \ni u$. The variational
approximation consists of simplifying the problem of
finding stationary solutions of~(\ref{eqn:u}) by
restricting the space of functions $\calD$
to a family of functions $u=U(x,A_1,\ldots,A_n)$ described
by a finite number of parameters $A_1,\ldots,A_n$. Such functions
span a finite dimensional subspace $\calD_A \subset \calD$. We denote
the restricted Lagrangian by $\calLeff:=\calL|_{\cal{D}_A}$.
Extrema for the effective Lagrangian satisfy the finite number
of constraints $\p \calLeff / \p A_i=0$ for $i=1,\ldots, n$.
Explicitly, these constraints can be written as
\ba
\frac{\p \calLeff}{\p A_i} = \int_\Omega \frac{\p U}{\p A_i} \left[ 
\frac{d}{dx}\frac{\p F}{\p U_x} - \frac{\p F}{\p U}
\right] \ dx & = & 0, \, i=1,\ldots, n. \nn
\ea

Suppose now that we allow evolution in time of the parameters $A_i(t)$.
Restricting~(\ref{eqn:u}) to $\calD_A$ we now have
\ba
\frac{\p F}{\p U} - \frac{d}{dx}\frac{\p F}{\p U_x} & = & 
\sum_{j=1}^n \frac{\p U}{\p A_j} \dot A_j
- f_{NV}\left(x,U,U_x,U_{xx}\right), \nn 
\ea
where $\dot A_j \equiv dA_j/dt$, so that for each $i$ we have
\ba
\frac{\p \calLeff}{\p A_i}&=& \int_\Omega 
\frac{\p U}{\p A_i} \left[
\frac{d}{dx}\frac{\p F}{\p U_x} - \frac{\p F}{\p U} \right]
\ dx, \nn\\
&=& \int_\Omega \frac{\p U}{\p A_i}\left[
f_{NV} - \sum_{j=1}^n\frac{\p U}{\p A_j} \dot A_j 
\right]\ dx.
\nn
\ea
Interchanging the integration and finite summation we obtain a set
of ordinary differential equations for the parameters $A_j$:
\ba
\sum_{j=1}^n M_{ij} \dot A_j = -\frac{\p \calLeff}{\p A_i}
-\int_\Omega \frac{\p U}{\p A_i}\ f_{NV}\ dx, \label{eqn:odes}
\ea
where the coupling matrix $M_{ij}=\int_\Omega \frac{\p U}{\p A_i}
\frac{\p U}{\p A_j} \ dx$ is clearly symmetric. We note that
the variational setup also shows that when $f_{NV}\equiv0$, $\calLeff$ is a non-increasing
function of time, since
\ba
\frac{d \calLeff}{dt} &=& \sum_{i=1}^n \dot A_i \frac{\p \calLeff}{\p A_i}\nn\\
&=& \int_\Omega \sum_{i=1}^n \dot A_i 
\frac{\p U}{\p A_i} \left[ 
\frac{d}{dx}\frac{\p F}{\p U_x} - \frac{\p F}{\p U}
\right] \ dx\nn\\ &=& -\int_\Omega \left( 
\sum_{i=1}^n \dot A_i \frac{\p U}{\p A_i}
\right)^2 \ dx \leq 0. \nn
\ea
We note that, equivalently, $\frac{d \calLeff}{dt}$ can be expressed
in terms of the coupling matrix $M$:
\ba
\frac{d \calLeff}{dt} = -
\sum_{i,j=1}^n \dot A_i M_{ij} \dot A_j. \nn
\ea

\section{Propagating fronts with spatial inhomogeneity}
\label{sec:fisher}

In this section we apply the general formulation from section~\ref{sec:dyn} to
the specific problem of a stable travelling front in a scalar
nonlinear diffusion equation with a bistable cubic nonlinearity. We introduce a
space-dependent factor multiplying the nonlinear term and investigate
the variation in front propagation speed that this causes.
Consider the PDE
\ba
u_t = u_{xx} + s(x)u(1-u)(u-a) = f_V(x,u,u_x,u_{xx}) , \label{eqn:pde}
\ea
having constant (and linearly stable) solutions $u(x,t)=1$ and $u(x,t)=0$,
and a linearly unstable solution $u(x,t)=a$. We take the
parameter $a$ to satisfy $0<a<1/2$ without loss of generality (the
dynamics for $1/2<a<1$ are equivalent after the change of variable
$u = 1-\tilde u$). 
We remark in passing that even though the PDE~(\ref{eqn:pde})
is not constant coefficient, it
is nevertheless variational since it can be written
in the form~(\ref{eqn:u}) - (\ref{eqn:fv}) with $f_{NV}\equiv0$ and
\ba
F & = & -\frac{1}{2}u_x^2 - s(x)\left( \frac{1}{4}u^4 
- \frac{1}{3}(1+a)u^3 + \frac{a}{2}u^2 \right). \nn
\ea

In the homogeneous case where \jhpd{$s(x)\equiv s_0$} is
constant there is a stable, and monotonically decreasing,
travelling front (or `kink') solution
\ba
u(x,t)= \left[ 1+\exp(\sqrt{s_0/2}(x-ct)) \right]^{-1}, \label{eqn:front}
\ea
with speed $c=\sqrt{s_0/2}(1-2a)>0$. We note that a
symmetrically-related monotonically increasing front also exists.

When we allow $s(x)$ to vary in space, we anticipate that if $s(x)$
varies slowly compared to the characteristic width of the front
$\sim \sqrt{2/s}$ then the solution will remain close to a monotonic
front of the form~(\ref{eqn:front}), for example being
well-approximated by an ansatz of the form
\ba
u(x,t)= \left[ 1+\exp((x-x_0(t))/w(t)) \right]^{-1}, \label{eqn:ansatz}
\ea
The method of section~\ref{sec:dyn} then enables us to deduce
ODEs for the parameters $x_0(t)$ and $w(t)$ characterising the
position and width of the travelling front as it passes, for example
over a region in which $s(x)$ varies.

To simplify notation, following Caputo and Sarels \cite{CS11},
we write our ansatz~(\ref{eqn:ansatz}) in the form
\ba
u(x,t)= U(z) = [1+\exp(z)]^{-1}, \qquad z=(x-x_0)/w. \nn
\ea
The ODEs~(\ref{eqn:odes}) for the evolution of the parameters $x_0$ and $w$
take the form
\ba
\dot x_0 \int_{-\infty}^\infty \frac{\p U}{\p x_0} \frac{\p U}{\p x_0} \ dx
&+& \dot w \int_{-\infty}^\infty \frac{\p U}{\p x_0} \frac{\p U}{\p w} \ dx\nn\\
& = & \int_{-\infty}^\infty \frac{\p U}{\p x_0} f_V \ dx, \label{eqn:x01} \\
\dot x_0 \int_{-\infty}^\infty \frac{\p U}{\p w} \frac{\p U}{\p x_0} \ dx
&+& \dot w \int_{-\infty}^\infty \frac{\p U}{\p w} \frac{\p U}{\p w} \ dx\nn\\
& = & \int_{-\infty}^\infty \frac{\p U}{\p w} f_V \ dx. \label{eqn:w1}
\ea
Compare~(\ref{eqn:x01}) - (\ref{eqn:w1}) with
(14)-(15) of \cite{CS11}. The integrals on the right-hand side evaluate to yield
\ba
\int_{-\infty}^\infty \left( \frac{\p U}{\p x_0} \right)^2 \ dx
& = & \frac{1}{w}\int_{-\infty}^\infty \frac{\e^{2z}}{(1+\e^z)^4} \ dz 
= \frac{1}{6w}, \nn \\
\int_{-\infty}^\infty \frac{\p U}{\p x_0} \frac{\p U}{\p w} \ dx
& = & \frac{1}{w}\int_{-\infty}^\infty \frac{z \e^{2z}}{(1+\e^z)^4} \ dz 
= 0, \label{eqn:crossterm} \\
\int_{-\infty}^\infty \frac{\p U}{\p w} \frac{\p U}{\p w} \ dx
& = & \frac{1}{w}\int_{-\infty}^\infty \frac{z^2 \e^{2z}}{(1+\e^z)^4} \ dz, \nn\\
&=& \frac{1}{3w}\left(\frac{\pi^2}{6} - 1 \right). \nn
\ea
That~(\ref{eqn:crossterm}) vanishes can be seen immediately since
the integrand is odd-symmetric. Hence in this case the coupling matrix
$M$ happens to be diagonal.
Now we turn to the right-hand sides of~(\ref{eqn:x01}) - (\ref{eqn:w1}).
The contributions from the diffusive term $u_{xx}$ in $f_V$ turn out to
be zero (since our ansatz $U(z)$ has zero slope as $x \ra \pm \infty$)
and $1/(12w^3)$, respectively. The nonlinear terms in $f_V$ make
contributions that depend on the (as yet unspecified) form of $s(x)$. We
obtain
\ba
\frac{\dot x_0}{6w}  &=&  \int_{-\infty}^\infty S\,
 \frac{\e^{2z}(1-a-a\e^z)}{(1+\e^z)^5} \ dz, \label{eqn:x02} \\
\frac{\dot w}{3w}\left( \frac{\pi^2}{6}-1 \right)
 &=&  \frac{1}{12w^3} + \frac{1}{w}\int_{-\infty}^\infty S\,
\frac{z\e^{2z}(1-a-a\e^z)}{(1+\e^z)^5} \ dz, \nn \\ & & \label{eqn:w2}
\ea
where $S=s(zw+x_0)$. We note that this form is similar to that obtained by \cite{CS11}; the difference
is  that they obtained a pair of ODEs for $x_0(t)$ and $w(t)$
in a more arbitrary fashion by multiplying~(\ref{eqn:pde}) by
$1$ and by $u$, respectively, before integrating over $\R$.
In this way Caputo and Sarels obtained the ODEs
\ba
\dot x_0 & = & w \int_{-\infty}^\infty S\,
 \frac{\e^{z}(1-a-a\e^z)}{(1+\e^z)^3} \ dz, \label{eqn:x-cs} \\
\dot w & = & \frac{1}{3w} - w \int_{-\infty}^\infty S\,
\frac{\e^z(1-\e^z)(1-a-a\e^z)}{(1+\e^z)^4} \ dz. \label{eqn:w-cs}
\ea
These are equations (19) in \cite{CS11}.

\section{Numerical comparisons}
\label{numr}

In this section we present a small number of comparisons between the
two reduced `collective coordinates' descriptions of the front
\jhpd{dynamics: the equations~(\ref{eqn:x02}) - (\ref{eqn:w2}) derived
here and~(\ref{eqn:x-cs}) - (\ref{eqn:w-cs}), derived in \cite{CS11}.}

\subsection{Slowly-varying heterogeneity $s(x)$}

An obvious consistency check of both pairs of ODEs is that they reduce to
the known exact front solution~(\ref{eqn:front}) when $s(x)$ is constant. Moreover,
this should be the asymptotic behaviour of the system when $s(x)$ varies only on
long lengthscales. In such a situation it is appropriate to replace $S=s(zw+x_0)$
in~(\ref{eqn:x02}) - (\ref{eqn:w-cs}) by $s(x_0)$. For our
\jhpd{formulation}~(\ref{eqn:x02}) - (\ref{eqn:w2}), \jhpd{after this
additional approximation} we obtain
\ba
\dot x_0 & = & \left( \frac{1}{2} - a \right) w s(x_0), \nn \\
\frac{\dot w}{3w}\left( \frac{\pi^2}{6} - 1 \right) 
& = & \frac{1}{24w^3}\left( 2-w^2 s(x_0) \right). \nn 
\ea
\jhpd{The second of these equations has a stable
equilibrium at $w=\sqrt{2/s(x_0)}$; this in turn implies} a value
$\dot x_0=(1-2a)\sqrt{s/2}$ for the
speed of the front. All these observations agree with
properties of the exact solution~(\ref{eqn:front}). The Caputo--Sarels
ODEs~(\ref{eqn:x-cs}) - (\ref{eqn:w-cs}) also reduce correctly in this
limit. 

\subsection{Wide defects}

\begin{figure}[!h]
\bc
\includegraphics[width=9.0cm]{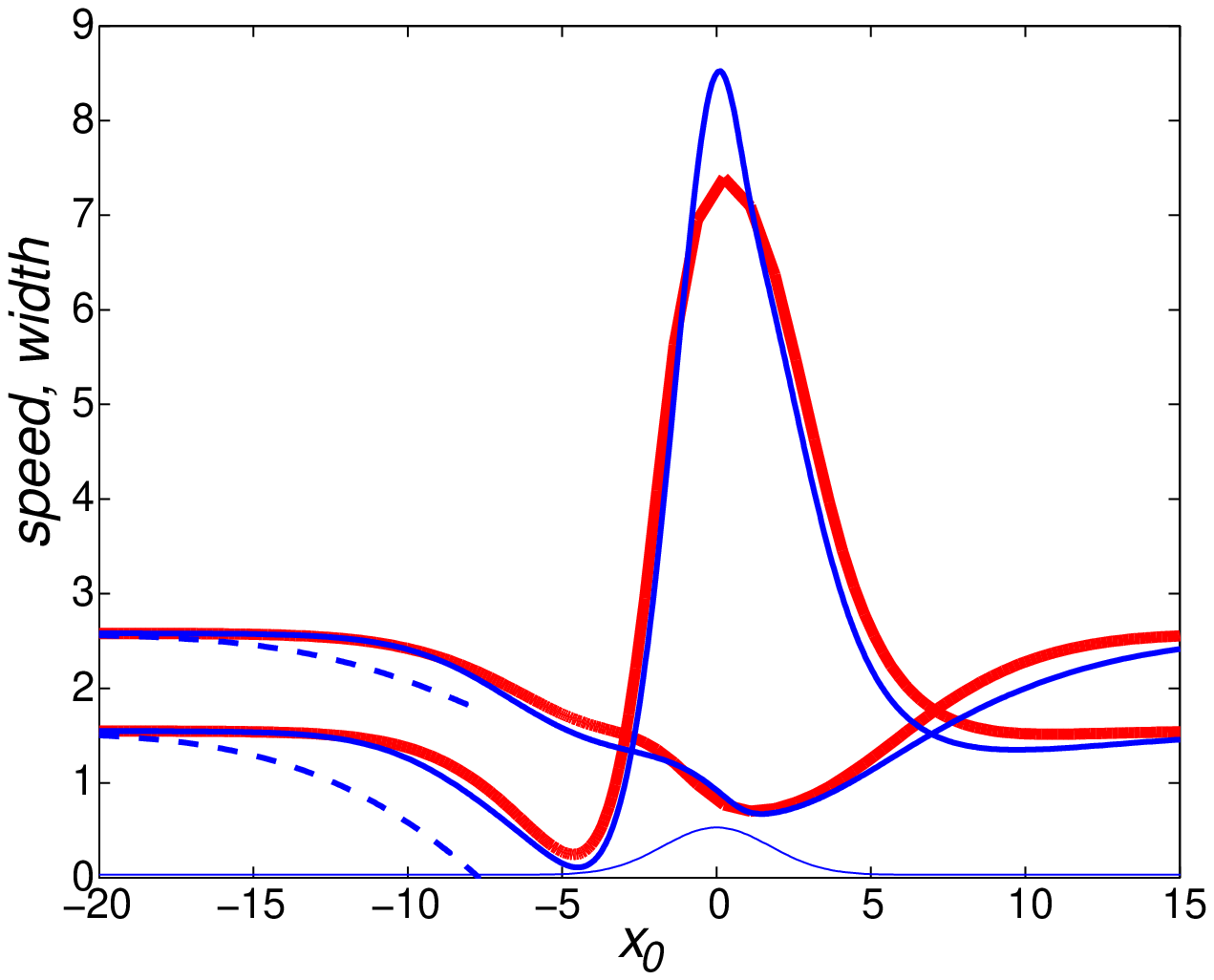}   
\includegraphics[width=9.0cm]{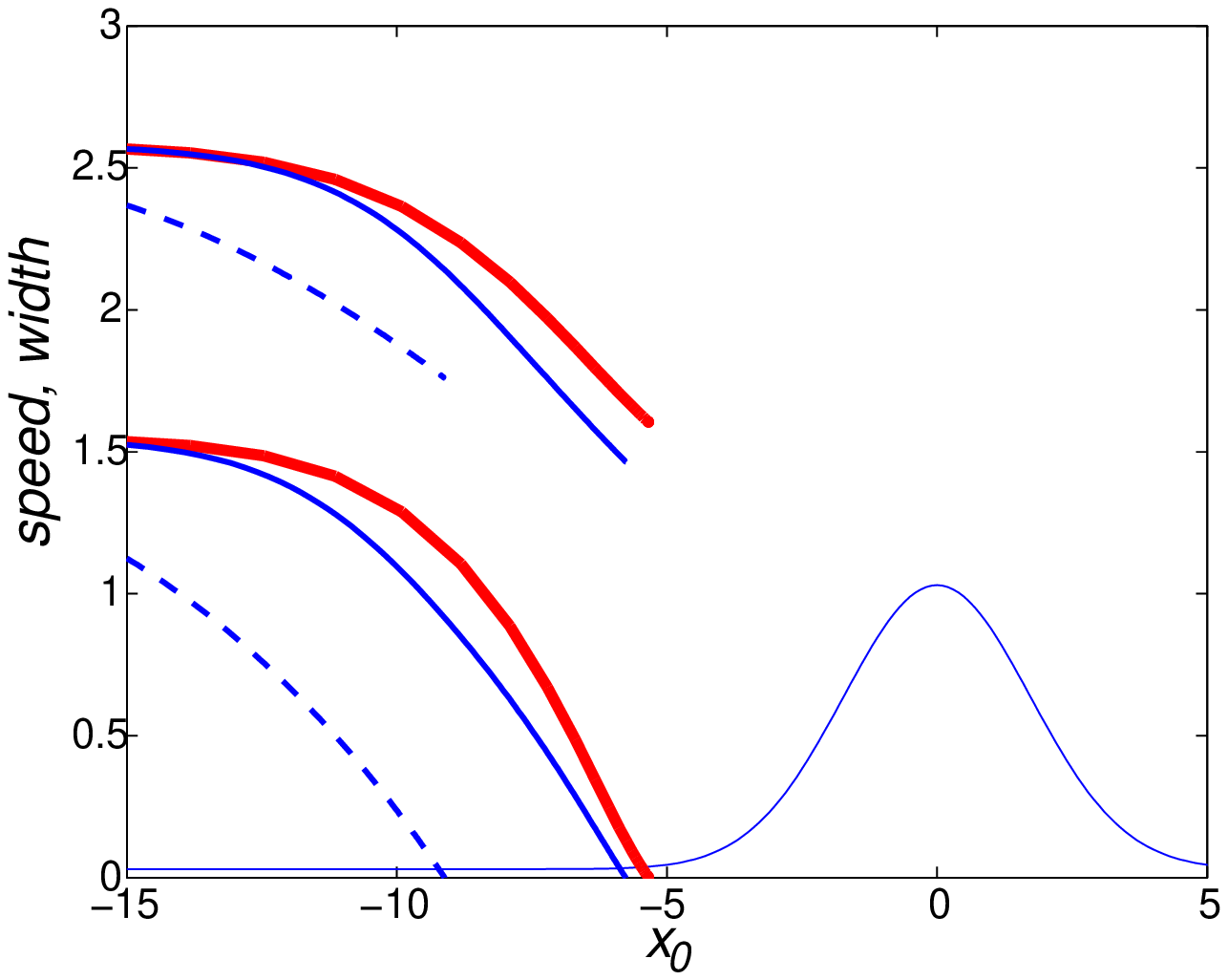}  
\caption{(Color online). Evolution of the width $w(t)$ and speed $\dot x_0(t)$
of a front \jhpdtwo{starting at $x_0=-40$ and} passing through a region
of varying $s(x)$, as a function of the position $x_0(t)$ of the centre of the
front, defined to be where $u(x_0)=1/2$.
Three sets of results are shown: best-fit to the numerical
solutions (thick solid red curves), the solution of the
ODEs~(\ref{eqn:x02}) - (\ref{eqn:w2}) (medium weight solid blue curves),
and the solution of the ODEs~(\ref{eqn:x-cs}) - (\ref{eqn:w-cs}) (dashed blue
curves). In each case the upper line is the width parameter $w(t)$ and the lower
is the speed $\dot x_0(t)$. Also shown is the form of the defect:
the plot shows $s(x)/10$ (thin blue curve).
Parameters are: $a=s_0=0.3$, $d=3$  and (a) $s_1=5$; (b) $s_1=10$.}
\label{fig:fig1}
\ec
\end{figure}

To \jhpd{investigate} the accuracy of the reduced equations for heterogeneities
$s(x)$ that are slowly-varying but not small amplitude, we consider
Gaussian and sinusoidal forms
\ba
s(x) & = & s_0 + s_1 \exp (-x^2/(2d)), \label{eqn:gaussian} \\
s(x) & = & s_0 + s_1 \sin(2\pi x/d). \label{eqn:sin}
\ea
Figure~\ref{fig:fig1} presents numerical comparisons between
the speed and width of the right-travelling
front solution, as estimated from a well-converged
numerical solution to the PDE~(\ref{eqn:pde}), and the two
approximations~(\ref{eqn:x02}) - (\ref{eqn:w2})
and~(\ref{eqn:x-cs}) - (\ref{eqn:w-cs}).
\jhpdtwo{We take the initial
condition to be the exact solution~(\ref{eqn:ansatz}) centred
at $x_0=-40$, far from the centre of the Gaussian inhomogeneity.}
We follow the method of
\cite{CS11} by, at a fixed point in time,
estimating the best-fit values of $x_0$ and $w$ via a least-squares fit
of the numerical solution to the kink solution~(\ref{eqn:ansatz}).
This procedure can be thought of as a projection in the
space of available functions.
\jhpdtwo{More precisely, we use the MATLAB routine \texttt{lsqcurvefit}
that implements an interior-reflective Newton method to solve the nonlinear least-squares problem of fitting the spatial form of the solution $u(x,t)$
at fixed time $t$ to the ansatz~(\ref{eqn:ansatz}). This produces
the least-squares optimal estimate of the parameters $x_0$ and $w$ at that time $t$.
The speed of the front $\dot x_0$ is subsequently estimated by taking
a centred second-order finite difference.}

Figure~\ref{fig:fig1}(a)
shows a typical solution in the case of a wide and
large-amplitude Gaussian defect, where
the front slows rapidly and narrows before accelerating rapidly to move
through the region around $x=0$ in which $s(x)$ is larger.
\jhpdtwo{The curves all agree closely near $x=-40$ and so for clarify we plot
only the range $-15 \leq x \leq 5$ where their divergence is most
obvious.}
The solution
of the reduced system~(\ref{eqn:x02}) - (\ref{eqn:w2}), shown as the 
solid blue curves, tracks the projection of the
numerical solution of the PDE (solid red curves) throughout the motion.
In contrast, the \jhpd{solution of the}
ODEs~(\ref{eqn:x-cs}) - (\ref{eqn:w-cs}), shown as dashed
blue curves, departs from the (numerical) values much further
away from the defect (certainly by the point $x=-10$).

Figure~\ref{fig:fig1}(b) illustrates the effect of increasing the height
of the defect further: the front speed drops to zero, \jhpd{i.e. the front
is now pinned near the defect}, and the front width
remains constant. As in figure~\ref{fig:fig1}(a) it is clear that the
ODEs~(\ref{eqn:x02}) - (\ref{eqn:w2}) provide a much more reliable guide
\jhpd{both to the dynamics of the system, and to the position at
which the front ultimately stops moving, as indicated by the lower curves
in the figure falling to zero.}

\begin{figure}[!h]
\bc
\hspace{-0.5cm}\includegraphics[width=9.0cm]{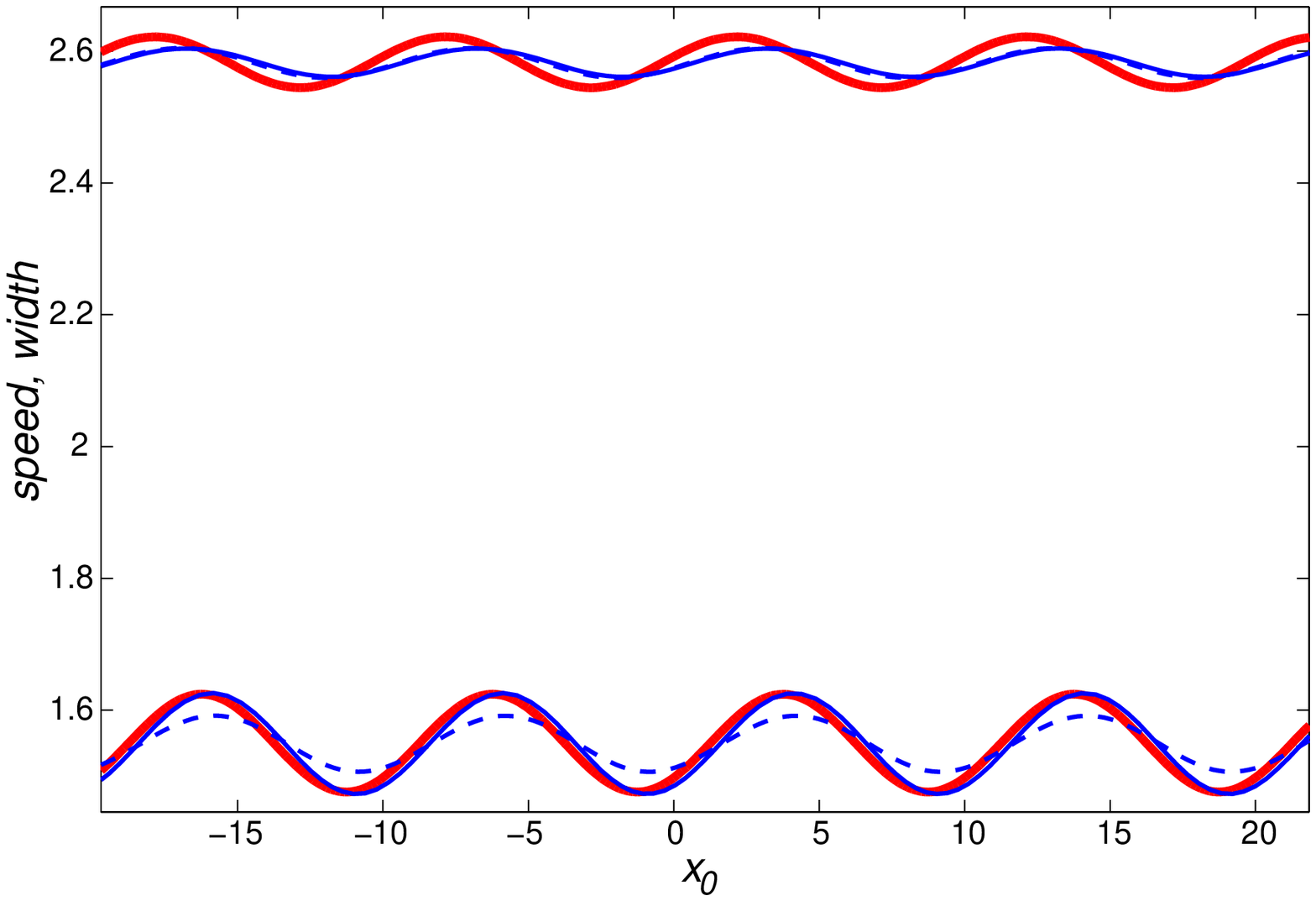}  

\hspace{-0.5cm}\includegraphics[width=9.0cm]{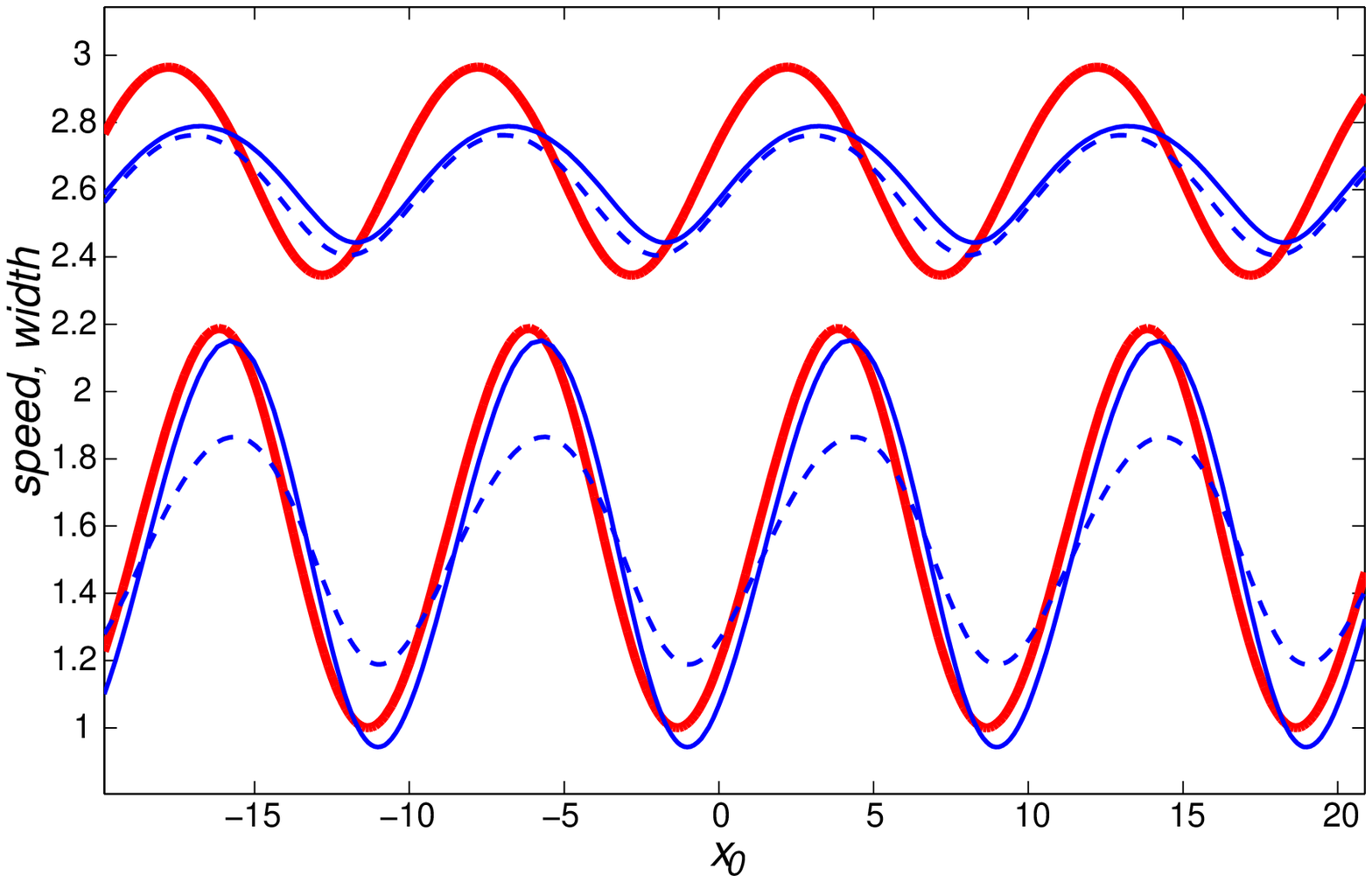}  
\caption{(Color online). Evolution of the width $w(t)$ and speed $\dot x_0(t)$
of a front subjected to a sinusoidal variation in $s(x)$,
as a function of the position $x_0(t)$ of the centre of the
front, defined to be where $u(x_0)=1/2$. \jhpdtwo{The initial
condition is~(\ref{eqn:ansatz}) centred at $x_0=-40$.}
\jhpd{In each plot we show
three sets of results:} best-fit to the numerical
solutions (thick solid red curves), the solution of the
ODEs~(\ref{eqn:x02}) - (\ref{eqn:w2}) (medium weight solid blue curves),
and the solution of the ODEs~(\ref{eqn:x-cs}) - (\ref{eqn:w-cs}) (dashed blue
curves). \jhpd{For each linestyle} the upper curve
is the width parameter $w(t)$ and the lower
is the speed $\dot x_0(t)$. Parameters are: $a=s_0=0.3$, $d=10$ and
(a) $s_1=0.025$; (b) $s_1=0.2$.}
\label{fig:sin}
\ec
\end{figure}

Figure~\ref{fig:sin} illustrates the accuracy of the approximations
when we take the sinusoidal form~(\ref{eqn:sin})
for the background, \jhpd{instead of the Gaussian}.
\jhpdtwo{For this numerical simulation we used the initial condition~(\ref{eqn:ansatz}) starting at $x_0=-40$. In this case the
sinusoidal inhomogeneity occupies the entire computational domain $-40 \leq x \leq 40$
so that in this case the curves depart immediately from each other as the computation begins. Over the range of $x$ shown in figure~\ref{fig:sin}
the computations have however settled into periodic fluctuations that
illustrate nature of the dynamics at long times.}

We observe that our variational method produces
much better agreement with numerical simulation for the speed of the front
\jhpd{(the lower curves) in both plots. Interestingly, the variation in front width
is not captured well by either collective coordinate approach,
even for small-amplitude variations in $s(x)$
as shown in figure~\ref{fig:sin}(a)}. It appears that the width of the
front takes far longer
to adjust to variations than the speed of the front does. In contrast,
figure~\ref{fig:sin}(b) indicates that the variational ODEs~(\ref{eqn:x02})
- (\ref{eqn:w2})
provide a good guide to the evolution of the speed even when the fluctuations
in $s(x)$ are very large: note that the fluctuations in $s(x)$ cover the range
$0.1 \leq s(x) \leq 0.5$ in this case.

\subsection{Narrow \jhpd{Gaussian} defects}

As the spatial variation in $s(x)$ becomes more rapid one would expect the
travelling front solution to the PDE~(\ref{eqn:pde}) to depart further from
the profile~(\ref{eqn:front}).

\begin{figure}[!h]
\bc
\includegraphics[width=9.0cm]{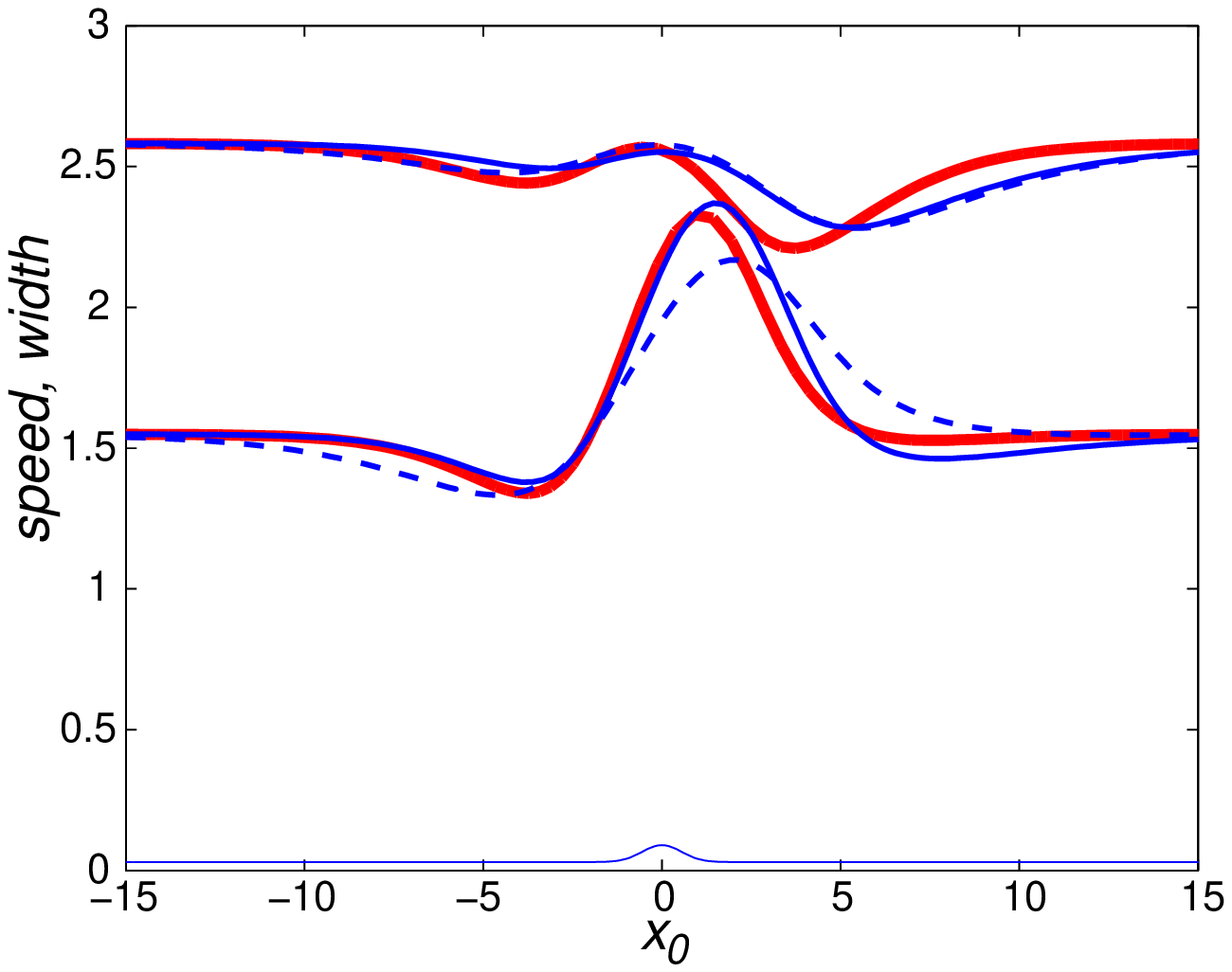}  
\includegraphics[width=9.0cm]{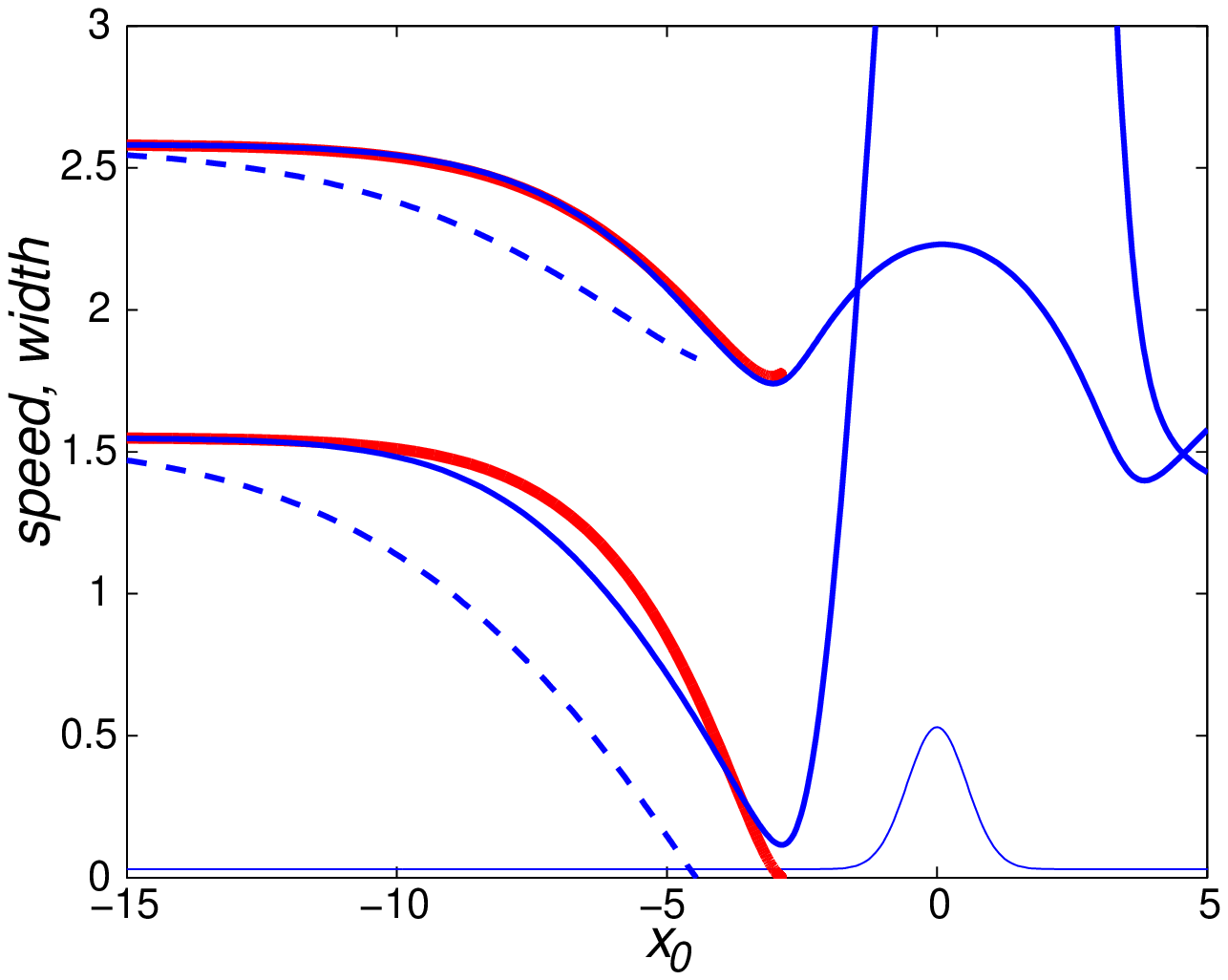}  
\caption{(Color online). Evolution of the width $w(t)$ and speed $\dot x_0(t)$
of a front, \jhpdtwo{with initial condition~(\ref{eqn:ansatz}), starting at $x_0=-40$, and} passing through a region
of varying $s(x)$, as for figure~\ref{fig:fig1}. Note that the speed
$\dot x_0$ is shown increased by a factor of 10 for clarity, and that
the defect is shown reduced by a factor of 10: $10\dot x_0$ and
$s(x)/10$ are plotted. Line colors and styles are as in figure~\ref{fig:fig1}.
Parameters: $a=s_0=d=0.3$ and (a) $s_1=0.6$; (b) $s_1=5$.}
\label{fig:fig2}
\ec
\end{figure}

Figure~\ref{fig:fig2} shows the results of both sets of reduced ODEs
in a typical case for a narrower defect. The parameter values used in
figure~\ref{fig:fig2} correspond to those in figure 10 of \cite{CS11}.
Figure~\ref{fig:fig2}(a) shows, as in figure~\ref{fig:fig1}, the width
and speed of fronts for the exact numerical solution (solid red curves),
the ODEs~(\ref{eqn:x-cs}) - (\ref{eqn:w-cs}) (dashed blue curves), and
the ODEs~(\ref{eqn:x02}) - (\ref{eqn:w2}) (solid blue curves). It is
interesting to observe that neither set of ODEs provides a particularly good
guide to the width of the front as it crosses the defect. The variation
in the front speed is captured much more closely by~(\ref{eqn:x02})
- (\ref{eqn:w2}) than by~(\ref{eqn:x-cs}) - (\ref{eqn:w-cs}).

Figure~\ref{fig:fig2}(b) presents a quantitative comparison for a case
with a narrow but much higher \jhpd{Gaussian} defect. Although the
the solution from the
ODEs~(\ref{eqn:x02}) - (\ref{eqn:w2}) (solid blue curves) lies much
closer to the true numerically-computed one than the approximation from
the ODEs~(\ref{eqn:x-cs}) - (\ref{eqn:w-cs}) (dashed blue curves), it
fails \jhpd{in this case} to predict that the front stops; instead it indicates a similar
slowing-down followed by an abrupt acceleration to that found for
the parameters of figure~\ref{fig:fig1}(a); \jhpd{clearly these parameter
values are close to the pinning threshold.}

\subsection{Estimates for front pinning}

In the case of a sufficiently strong and sufficiently spatially
localised unimodal inhomogeneity, for example the
Gaussian defect~(\ref{eqn:gaussian}), we expect that \jhpdtwo{the}
behaviour illustrated in figure~\ref{fig:fig1}(b) is typical: the front
width and position tend to constant values and the front is pinned by the
inhomogeneity. Although the integrals in~(\ref{eqn:x02}) - (\ref{eqn:w2})
are not usually available analytically, progress can be made in the special
case of a $\delta$-function: $s(x)=\alpha+\beta \delta(x)$. For this
\jhpd{point-like inhomogeneity we pose the following question: how large
must $\beta$ be} in order to pin the front, and what is the resulting
position of the front, when pinned? Substituting
\jhpd{the point-like} ansatz into~(\ref{eqn:x02}) - (\ref{eqn:w2}) we obtain
\ba
\frac{\dot x_0}{6w} & = & \frac{\beta}{w} \frac{\e^{-2x_0/w}\left( 
1-a-a\e^{-x_0/w} \right)}{(1+\e^{-x_0/w})^5} 
\nn \\ & & + \frac{\alpha}{12}(1-2a), \label{eqn:x0delta} \\
\frac{\dot w}{3w}\left( \frac{\pi^2}{6} - 1 \right) 
& = &  \frac{1}{12 w^3} - \frac{\alpha}{24w} \nn\\ 
& & - \frac{\beta x_0}{w^3}
\frac{\e^{-2x_0/w}\left( 1-a-a\e^{-x_0/w}\right)}{(1+\e^{-x_0/w})^5}. 
\label{eqn:wdelta}
\ea
For a pinned front we require $\dot x_0=\dot w=0$. Looking for equilibria
of~(\ref{eqn:x0delta}) - (\ref{eqn:wdelta}) we write~(\ref{eqn:x0delta}) in
the form
\ba
g(x_0/w) &=&-\frac{\alpha w}{12 \beta}(1-2a) \label{eqn:g}
\ea
\jhpd{where we define
\ba
g(x_0/w) &:=&  \frac{\e^{-2x_0/w}\left( 
1-a-a\e^{-x_0/w} \right)}{(1+\e^{-x_0/w})^5}. \nn\\
\ea}
We may now estimate the critical (i.e. smallest positive) value of $\beta$
required to pin the front by minimising $g(x_0/w)$ over its argument. The
relevant turning point of the two available
is the one at the more negative value of $z\equiv x_0/w$. We obtain
$z=z_{min}=\log((3-r)/[4(1-a)])$ where $r=\sqrt{16a(a-1)+9}$. Hence
\ba
g(z_{min}) & = & \frac{16(3-r-4a)(3-r)^2(1-a)^3}{(7-r-4a)^5}. \nn
\ea
Substituting \jhpd{$x_0/w=z_{min}$}, $g=g(z_{min})$, and $\beta=\beta_c$
into~(\ref{eqn:wdelta}) yields
\ba
\beta_c z_{min}g(z_{min}) = \frac{1}{12w}-\frac{\alpha w}{24}, \nn
\ea
which can be solved for $w$ to give
\ba
w_{min} & = & \left( \frac{2/\alpha}{1-2(1-2a)z_{min}} \right)^{1/2}. \nn
\ea
We can now write down the corresponding estimate of the pinning position $x_0$
since $x_{0}^{min}= w_{min} z_{min}$. Moreover, the critical value $\beta_c$
can be estimated using the relation~(\ref{eqn:g}) which was not directly used
in the last few lines above. We obtain the explicit estimates
\ba
\beta_c & = & \left( \frac{\alpha/2}{1-2(1-2a)z_{min}} \right)^{1/2}\times\nn\\
&&\frac{(1-2a)(7-r-4a)^5}{96(4a+r-3)(3-r)^2(1-a)^3}, \label{eqn:betac} \\
x_{0}^{min} & = & \left( \frac{2/\alpha}{1-2(1-2a)z_{min}} \right)^{1/2} z_{min}
\label{eqn:x0min}
\ea
where $z_{min}\equiv\log((3-r)/[4(1-a)])$ and $r=\sqrt{16a(a-1)+9}$
\jhpd{depend only on the PDE parameter $a$}.
Similar estimates were obtained by Caputo and Sarels~\cite{CS11} for
their ODE system, \jhpd{see their equations (31) and (32).}

\begin{figure}[!h]
\bc
\includegraphics[width=9.0cm]{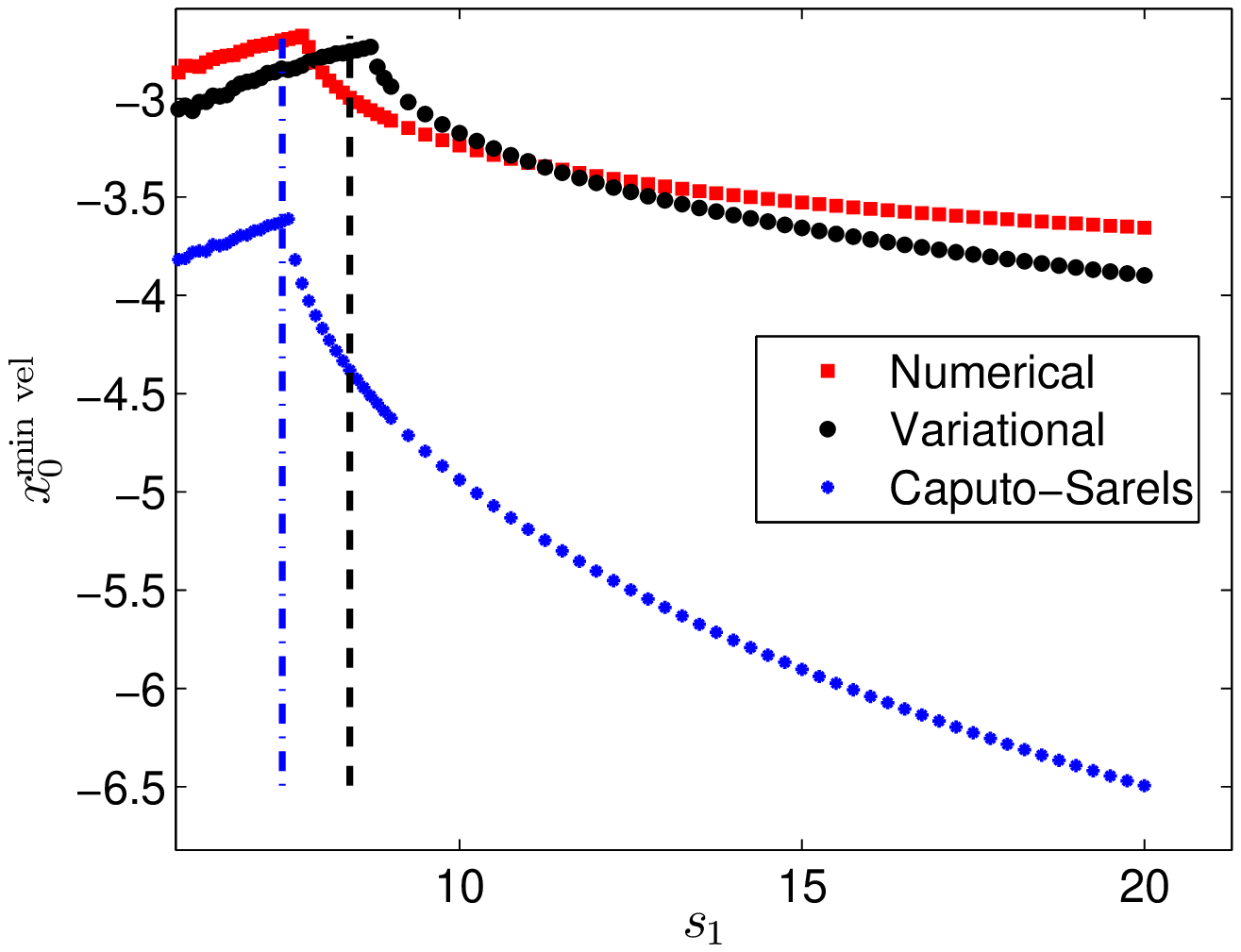}  
\includegraphics[width=9.0cm]{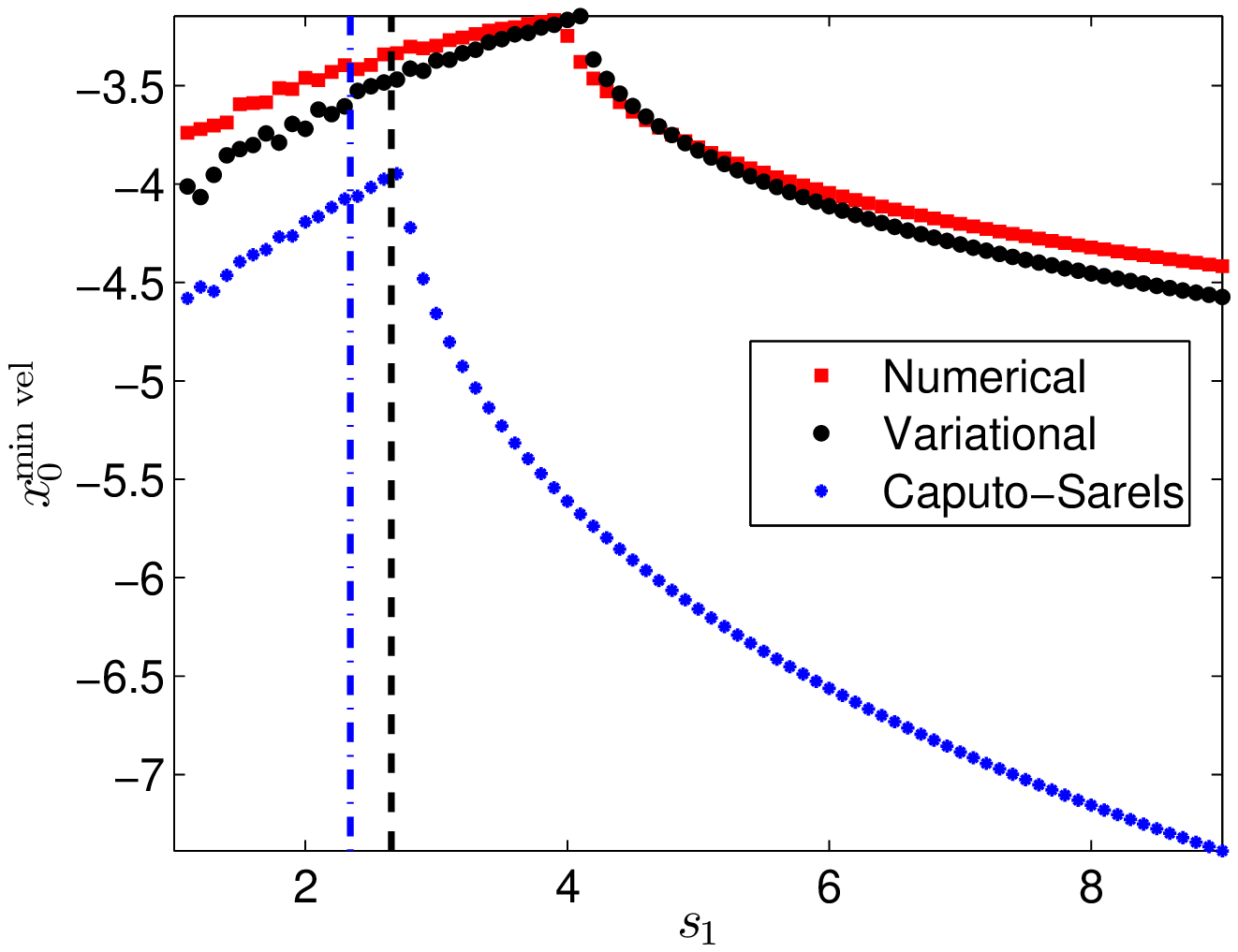}  
\caption{(Color online). Location of the minimum value of $\dot x_0$ as a function of
$s_1$ \jhpdtwo{for the Gaussian inhomogeneity~(\ref{eqn:gaussian})}.
Where the data \jhpd{values are increasing with increasing $s_1$}
the front is not pinned, but attains
a minimum velocity at the point shown. After the \jhpd{maximum data
value, in the range where the data are decreasing as $s_1$ increases},
the front is pinned (i.e. attains zero velocity) at the
point shown. Data are shown for numerical simulations (squares, red),
the ODEs~(\ref{eqn:x02}) - (\ref{eqn:w2}) (circles, \jhpd{black}) and the \jhpd{ODEs
(\ref{eqn:x-cs}) - (\ref{eqn:w-cs}) derived by}
\cite{CS11} (asterisks, blue). The vertical lines indicate the
estimates of the critical (i.e. lowest) $s_1=\beta_c/\sqrt{2\pi d}$ at which
the front is pinned. The variational~(\ref{eqn:betac}) and
Caputo--Sarels estimates are shown as the black dashed and blue dot-dashed
lines, respectively.
Parameters: $a=s_0=0.3$ and (a) $d=0.1$; (b) $d=1$.}
\label{fig:fig3}
\ec
\end{figure}

\jhpd{Figure~\ref{fig:fig3} presents comparisons of the numerically-determined
location $x_0^{\mathrm{min} \ \mathrm{vel}}$ at which a front attains its
minimum velocity while moving over the defect. For large enough $s_1$, above
the value at which $x_0^{\mathrm{min} \ \mathrm{vel}}$ achieves its maximum,
the fronts are pinned and come to rest at the locations indicated.
The results of the ODEs~(\ref{eqn:x02}) - (\ref{eqn:w2}) and
the estimates provided by~(\ref{eqn:betac}) are compared with numerical values.
The figure also indicates, with the vertical lines, the
estimates of the minimum value of $s_1$ required in order to achieve pinning.
The locations of the vertical lines should be compared to the locations of the
maxima in the curves traced out by the (red) squares, (black) circles and (blue)
asterisks. Note that the black vertical dashed line obtained from the analytical
estimation using the point-like $\delta$-function ansatz is close to the maxima
in the red and black curves in the upper part of figure\ref{fig:fig3}, for which
the defect is more sharply localised than in the lower part of the figure.} 

\jhpd{More precisely, in} order to compare the
results for a Gaussian~(\ref{eqn:gaussian}) with the $\delta$-function
form, we equate the
\jhpd{integrals of $s(x)$ in the two cases. This implies the correspondence
$\beta=s_1 \sqrt{2\pi d}$.}
In both parts of figure~\ref{fig:fig3} we observe that the variational
ODEs are a substantially better guide to the pinning location than
that given in \cite{CS11}, although both give qualitatively a correct
description of the behaviour. The $\delta$-function approximation
is, as one might expect, substantially closer in figure~\ref{fig:fig3}(a)
for which $d=0.1$ and so the inhomogeneity occurs over a length-scale
less than the front width and much more like a $\delta$-function.
For larger $d$ (figure~\ref{fig:fig3}b), the value of $s_1$ needed to
pin the front decreases substantially. The values of $x_0$ at
which the front pins do not change significantly between
figure~\ref{fig:fig3}(a) and (b); this agrees with the
expression~(\ref{eqn:x0min}) for $x_0^{min}$ which depends only on the
parameters $a$ from the PDE and $\alpha$ which is the limiting
`far-field' value of $s(x)$.

\section{Conclusion}
\label{sec:discussion}

In this paper we have discussed the relation between the variational approximation and collective coordinate methods.
We showed that the variational method implies a natural choice of functions to use in deriving a reduced
description of the dynamics in terms of collective
coordinates.
\jhpdtwo{Our method also extends in a natural way to non-variational problems; in this work this extension is not used, but it points the way to a unified
approach for both variational and non-variational situations that we believe
is a scientifically valuable observation.}
For the specific problem considered in section~\ref{sec:fisher}, this formulation of the collective
coordinate reduction yielded a much more accurate approximation than an existing treatment in the
literature \cite{CS11}.

Many questions arise which we will return to in future work. Firstly, we have not provided any
systematic approach to estimating how accurate we might expect this approach to be. Bounds on the accuracy
that one might expect would be extremely useful. 
\HS{Error bounds for variational approximations for Lagrangian systems having equilibrium or travelling wave solutions have been given in \cite{kaup07,chon12}. To our knowledge, the only existing analyses for the fully dynamic case similarly to that we work with here only consider slowly varying position of the front, i.e.\ $x_0(t)$, (see, e.g., \cite{vaku91,vaku97,ei02,ei02_2}).} 
Secondly, it is questionable whether
or not the method than we outline above maintains its accuracy when there are different length scales present in the problem, since previous work \cite{sanc98} has indicated that competition
between different length scales may strongly influence the applicability of collective
coordinate methods \cite{sanc98}. One specific example of this problem is the subcritical (and variational)
Swift--Hohenberg which is well-known to admit spatially localised equilibrium, travelling and time-periodically
pulsating states that have spatially oscillating tails \cite{D10,BD12}.
Inspired by the results of multiple-scales asymptotics, variational formulations of the problem
have enabled the development of good approximations to the dynamics of the localised states \cite{MS11,SM11}.
One natural extension of this work would be to analyse the interaction of such localised states with background
inhomogeneities of the kind discussed in this paper.

\begin{acknowledgments}
JHPD acknowledges support from the Royal Society through a University Research Fellowship.
\end{acknowledgments}

\end{document}